\documentclass[%
 reprint,
superscriptaddress,
 amsmath,amssymb,
 aps,
 prl,
]{revtex4-2}

\usepackage{graphicx}
\usepackage{dcolumn}
\usepackage{bm}
\usepackage{mathrsfs}
\usepackage{xcolor}
\usepackage{dblfloatfix}

\begin{document}

\title{Ultraslow vorticity modes in colloidal fluids}

\author{Satyam~Raheja}
\thanks{These authors contributed equally to this work.}
\affiliation{Department of Physics, Indian Institute of Technology Gandhinagar, Palaj, Gandhinagar, Gujarat, India.}

\author{Noman~Hanif~Barbhuiya}
\thanks{These authors contributed equally to this work.}
\affiliation{Department of Physics, Indian Institute of Technology Gandhinagar, Palaj, Gandhinagar, Gujarat, India.}

\author{Tsvi~Tlusty}
\affiliation{Department of Physics and Astronomy, Ulsan National Institute of Science and Technology, Ulsan, Republic of Korea.}

\author{Chandan~K.~Mishra}
\email{chandan.mishra@iitgn.ac.in}
\affiliation{Department of Physics, Indian Institute of Technology Gandhinagar, Palaj, Gandhinagar, Gujarat, India.}

\date{\today}

\begin{abstract}
In the conventional unsteady-Stokes picture, vorticity generated by a transient disturbance relaxes rapidly through viscous diffusion once the forcing ceases. This view presumes a quiescent solvent. In colloidal and biological fluids, thermal motion continually perturbs the surrounding fluid, raising a natural question: can equilibrium fluctuations, mediated by hydrodynamic coupling, sustain vorticity over mesoscopic scales and times? Here, using high-speed confocal microscopy of moderately dense colloidal suspensions, we reconstruct the conditional mean hydrodynamic response to Brownian displacements of prescribed magnitude and direction. We find a long-lived, spatially structured vorticity field that decays far more slowly than predicted by unforced unsteady-Stokes dynamics. A coarse-grained phenomenological model based on fractional Brownian motion, constrained by independently measured single-particle subdiffusion, quantitatively captures its temporal and spatial decay. These findings reveal a field-level manifestation of hydrodynamic memory in fluctuating low-Reynolds-number flows, with implications for transport and collective dynamics in complex fluids.
\end{abstract}

\maketitle
\paragraph{Introduction---}Vorticity is both a hallmark and driver of turbulent flows at high Reynolds number (Re), where inertia and advection set the dynamics~\cite{batchelor2000introduction, frisch1995turbulence, sreenivasan1999fluid, eyink2006onsager, holland1984dynamics, buaria2020self, buaria2024twisting, de2024pattern}. Its status in the opposite, Stokesian limit of low Re is much less settled, even in the most basic textbook system of passive colloidal particles driven by thermal agitation~\cite{guyon2015physical, Happel1983, cuevas2006vorticity}. Here, inertial nonlinearities fade away, yet every Brownian particle executes random motions that continually stir the fluid and generate time-varying flow fields~\cite{landau2013fluid, keblinski2006hydrodynamic, fares2024observation}. On general grounds, one expects these flows to carry local vorticity. But despite significant experimental and theoretical progress, the nature of such vorticity in colloidal suspensions remains essentially unknown.

In theory and computation, there has been much progress in stochastic hydrodynamics and in long-time velocity auto-correlation function (VACF) of hydrodynamic memory~\cite{lukic2005direct, Franosch2011, huang2011direct, Kheifets2014, huang2015effect, mandal2019persistent}. In experiment, high-precision measurements show that a Brownian particle remains dynamically coupled to the surrounding fluid through hydrodynamic memory, producing the familiar $t^{-3/2}$ long-time tail in the VACF~\cite{Franosch2011, Kheifets2014}. This algebraic decay reflects relaxation of momentum over a sub-microsecond characteristic time, $\tau_p \sim m/\gamma \sim 10^{-7}~\mathrm{s}$ (for a colloidal particle of mass $m$ and drag $\gamma$), naturally pointing to a vorticity cloud that feeds back on the particle's motion.

Yet the solvent vorticity field itself$-$the actual carrier of this memory$-$has remained largely invisible, mostly because the fast, fine-scale vortical structures are extremely difficult to resolve in real time. Thus, it has been established that Brownian particle motion generates transient vorticity in the surrounding fluid, but an open question remained: Do these Brownian-motion--driven swirls always die as fleeting transients or perhaps can collectively interact and accumulate correlated vorticity over much longer spatiotemporal scales? In moderately concentrated suspensions, where many-body hydrodynamic interactions are non-negligible, the latter becomes physically plausible: many-body coupling can collectively amplify the underlying single-particle memory, making the vortical response stronger and more persistent.

\begin{figure}[htb!]
    \includegraphics[width=\linewidth]{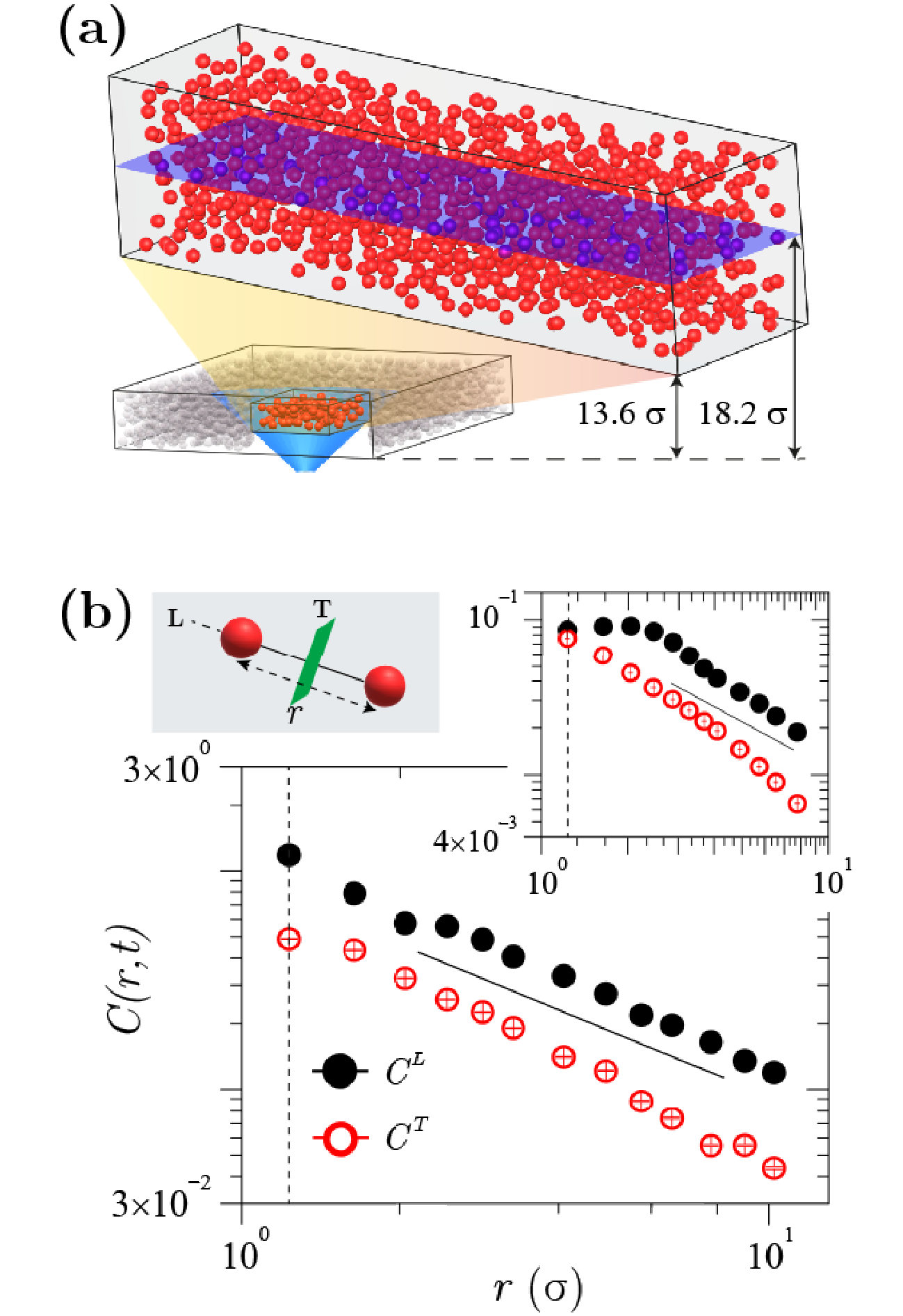}
    \caption{\textbf{Experimental set-up and hydrodynamic correlations.}
        (\textbf{a}) A representative rendering of three-dimensional (3D) experimental chamber at a particle volume fraction of $\Phi \sim 0.12$. The particles in the field-of-view are rendered as red spheres. The inset shows an enlarged view of the experimental region from which both 3D data (from the entire volume) and 2D data (at the blue mid-plane) are acquired.
        (\textbf{b}) Hydrodynamic longitudinal $C^L(r,t)$ and transverse $C^T(r,t)$ velocity correlations on a log-log scale for 2D data (main plot, $t=4.67$~s) and 3D data (inset, $t=1.72$~s); units are $\sigma^2\,\mathrm{s}^{-2}$. The inset shows the $L$ and $T$ directions for a representative pair of particles. The vertical dashed lines at $r \sim 1.1 \sigma$ indicate the range of the inter-particle potential. Solid $1/r$ lines are shown as guides to the eye. Error bars represent the standard error of the mean calculated from time- and ensemble-averaging.}
    \label{fig:Figure1}
\end{figure}

Here, we address this question by combining high-speed, single-particle--resolved confocal microscopy (video~S1 \cite{supplemental}) with a complementary phenomenological model. We track individual colloids (diameter $\sigma \sim 2.2~\mu\mathrm{m}$) within a moderately dense suspension [Fig.~\ref{fig:Figure1}(a); End Matter for Materials and Methods]. This resolution enables a clean decomposition: each tracked particle serves as a localized source of fluid disturbance, while all others act as passive tracers sensing the hydrodynamic response. Importantly, a particle's net displacement $\mathbf{r}(t_0+t)-\mathbf{r}(t_0)$ over a short interval $t$ does not correspond to a deterministic trajectory but rather to an ensemble of stochastic paths connecting the two positions (Supplemental Fig. S1 \cite{supplemental}). By averaging over all source--tracer pairs, we extract the conditioned ensemble-averaged hydrodynamic flow field associated with a prescribed finite-lag Brownian displacement, thereby reconstructing the solvent's vorticity-bearing response conditioned on stochastic particle motion.

\paragraph{Hydrodynamic correlations---}To quantify the correlated motion between particles arising from hydrodynamic interactions, we compute the particle--particle velocity correlation function, shown in Fig.~\ref{fig:Figure1}(b) for a two-dimensional (2D) cross-section (main panel) extracted at the mid-plane of the bulk three-dimensional (3D) sample (inset). For each particle pair, we define a pair-centric coordinate frame with two orthogonal axes: the longitudinal direction $L$, along the line connecting their centers, and the transverse direction $T$, perpendicular to it [Fig.~\ref{fig:Figure1}(b), top left]. The longitudinal hydrodynamic velocity correlation for a pair separated by $r$ at initial time $t_0$ is $C^{L}(r, t) = \langle \mathbf{u}_i^{L}(t_0,t) \cdot \mathbf{u}_j^{L}(t_0,t) \rangle_{r, t_0}$, where the average velocity of the $i^{\mathrm{th}}$ particle in the pair is given by $\mathbf{u}_i(t_0,t) = [\mathbf{r}_i(t_0 + t) - \mathbf{r}_i(t_0)]/t$. Its longitudinal component is $\mathbf{u}_i^{L}(t_0,t) = [\mathbf{u}_i(t_0,t) \cdot \hat{\mathbf{r}}_{ij}]\,\hat{\mathbf{r}}_{ij}$, where $\hat{\mathbf{r}}_{ij}$ is the unit vector along $L$ at $t_0$. The transverse correlation, $C^{T}(r, t)$, is defined analogously. These correlations encode how the motion of one Brownian particle is hydrodynamically correlated with the motion of its neighbors through the underlying flow field, providing a direct probe of the solvent's vorticity and its spatial structure.

As expected and shown in Fig.~\ref{fig:Figure1}(b), both $C^L$ and $C^T$ remain positive well beyond particle-sized separations ($r > \sigma$), demonstrating that the motion of one particle entrains its neighbors through hydrodynamic interactions~\cite{Segre2001}. The comparatively weaker transverse response reflects the anisotropy inherent in this entrainment. At larger separations, the correlations decay algebraically with $r$, in a manner reminiscent of the spatial structure of a Stokeslet field~\cite{Kim2005}. This suggests that these long-ranged couplings are dominated by fluid-mediated momentum transport rather than direct particle contacts.

\begin{figure}[htb!]
    \includegraphics[width=\linewidth]{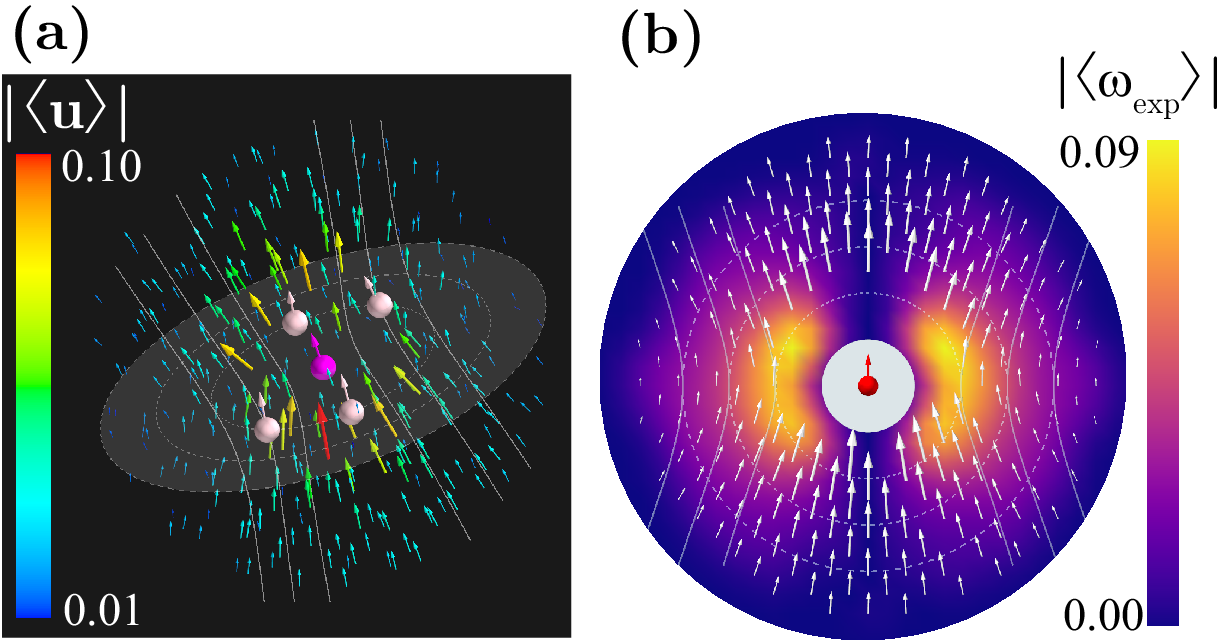}
    \caption{\textbf{Particle velocity and vorticity fields from 3D and 2D experimental data.}
        Velocity field (shown by arrows) generated by a moving source particle (at the origin), reconstructed from the average velocities of neighboring particles at each bin for (\textbf{a}) 3D experimental data ($t=1.72$~s) and (\textbf{b}) 2D experimental data ($t=0.05$~s). In both panels, the length and orientation of the arrows indicate the magnitude and direction of the time- and ensemble-averaged particle velocity. The gray solid curves (streamlines) represent the best-fit steady Stokeslet flow. Dashed concentric circles are drawn at $r=2\,\sigma$, $3\,\sigma$ and $4\,\sigma$ from the source particle. Four representative neighboring particles are shown in (\textbf{a}) at a distance of $r=2\,\sigma$, and velocity vectors are color-coded by magnitude. A light gray plane represents the equatorial plane. In (\textbf{b}), the background is color-coded by the local curl (vorticity). Vectors originating within $r=2\,\sigma$ of the source particle are not shown for visual clarity.}
    \label{fig:Figure2}
\end{figure}

\paragraph{Vorticity reconstruction---}We also reconstruct the conditioned ensemble-averaged velocity field by fixing the source particle at the origin, aligning its displacement along a prescribed direction, and treating all other particles as tracers of the induced hydrodynamic field (see End matter for Materials and Methods). By construction, this conditional average isolates the mean solvent response to a Brownian displacement of given sign and does not vanish by symmetry. As shown in Fig.~\ref{fig:Figure2}(a), neighboring particles are entrained and advected with the source particle, generating a flow pattern that closely mirrors the streamlines of a Stokeslet~\cite{Kim2005} (see Supplemental Note S1 \cite{supplemental}). Although a finite vorticity is evident in the 3D data (transverse plane), we employ faster 2D video microscopy experiments with better statistical confidence to quantitatively probe the temporal evolution of vorticity. We define the ensemble-averaged vorticity as $\langle \pmb{\omega}_{\mathrm{exp}} \rangle = \nabla \times \langle \mathbf{u} \rangle$, computed from the tracer-particle velocity field [Fig.~\ref{fig:Figure2}(b); video~S2 and Supplemental Fig.~S2 \cite{supplemental}]. Its statistical robustness is confirmed by subsampling analysis detailed in the Supplemental Material (Supplemental Fig.~S3) \cite{supplemental}.

The measured vorticity field deviates sharply from the rapidly decaying response predicted for an unforced, unsteady Stokes flow. Instead, vorticity persists over timescales far exceeding those set by solvent momentum diffusion, implying that fluctuations arising from coupled particle motions are not simply averaged away but accumulate through hydrodynamic interactions. Thus, vorticity in this Brownian suspension is not a fleeting by-product of individual particle displacements; it exhibits pronounced spatiotemporal correlations that appear to be sustained by the thermal fluctuations of the suspension itself.

\begin{figure}[htb!]
    \includegraphics[width=\linewidth]{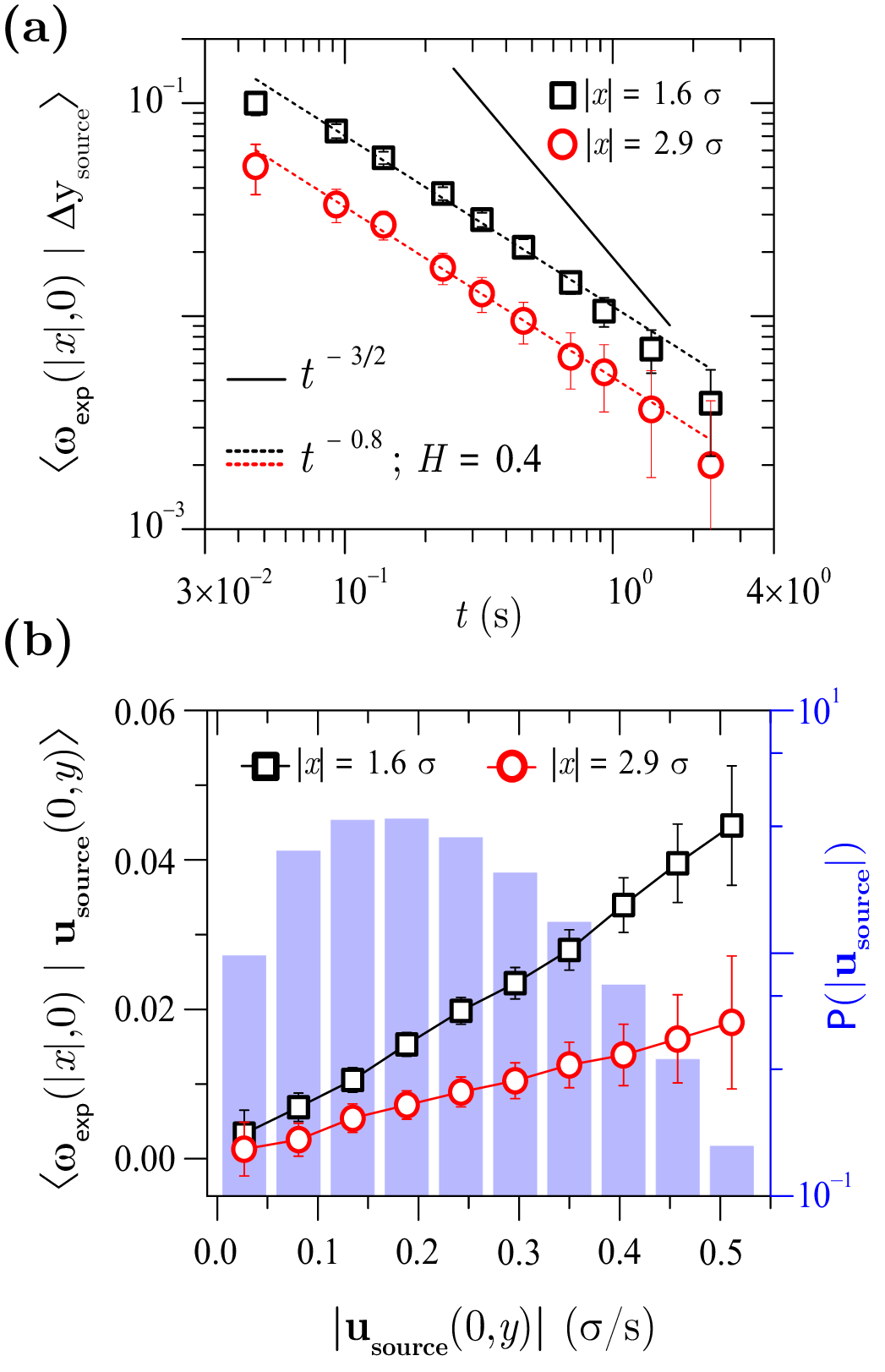}
    \caption{\textbf{Comparison of experimental vorticity decay with phenomenological model and its origin.}
        (\textbf{a}) The magnitude of experimentally measured vorticity (data points) is shown as a function of lag time $t$, with a fixed displacement constraint of $0.10\,\sigma \leq \Delta y_{\mathrm{source}} \leq 0.15\,\sigma$. Experimentally measured decay profiles are best captured by the phenomenological model incorporating anti-persistent correlations (Hurst exponent $H=0.4$; red and black short-dashed lines). Decay for an unforced unsteady Stokeslet (black solid line) is also shown.
        (\textbf{b}) Double $y$-axis plot comparing the magnitude of the induced vorticity (left axis, points) with the velocity of the source particle that generates it for a lag time of $t=0.93$~s. The probability density of the velocity of the reference particle is also shown as a histogram (right axis). In both sub-figures, the induced vorticity is shown at two positions, $|x|=1.6\sigma$ and $2.9\,\sigma$, along the transverse direction to the source particle, which is oriented along the $y$-axis (see End Matter for Materials and Methods). Error bars represent the standard error of the mean calculated from time- and ensemble-averaging.}
    \label{fig:Figure3}
\end{figure}

\paragraph{Ultraslow decay---}To elucidate the origin and temporal persistence of these thermally-induced flows, we quantify the temporal evolution $\langle \pmb{\omega}_{\mathrm{exp}}(r, t) \rangle$. To do this, we condition the position and displacement of the source particle, $\Delta y_{\mathrm{source}}\,\hat{\mathbf{y}}$, at the origin and along the $y$-axis and then measure $\langle \pmb{\omega}_{\mathrm{exp}}(r, t) \rangle$ at various pair separations $r$ along the transverse direction [Fig.~\ref{fig:Figure3}(a)]. At a fixed $r$, $\langle \pmb{\omega}_{\mathrm{exp}}(r, t) \rangle$ decays as a power law, $\langle \pmb{\omega}_{\mathrm{exp}}(r, t) \rangle \propto t^{-0.8}$. This ultraslow decay stands in stark contrast to the $t^{-3/2}$ behavior predicted by the unsteady Stokes equation, $\partial_t \pmb{\omega} = \nu \nabla^2 \pmb{\omega}$, which describes purely diffusive transport of vorticity. Here, $\nu$ denotes the kinematic viscosity, which also sets the diffusion constant for vorticity. Since the suspension is density- and index-matched, $\nu = \mu/\rho$, where $\mu$ is the dynamic viscosity and $\rho$ is the density of the particles and fluid. 

The significant deviations from the classical diffusive picture suggest that vorticity does not merely relax from isolated perturbations but is influenced by continuous momentum exchange between Brownian particles and the solvent. Such fluctuation-driven regeneration of vorticity effectively prolongs hydrodynamic memory far beyond microscopic timescales, suggesting a possible mechanism through which thermal fluctuations organize flow at mesoscopic scales in moderately dense suspensions.

To directly link thermal motion to the generation of vorticity, we exploit the statistical distribution of Brownian displacements $\Delta y_\mathrm{source} \, \hat{\textbf{y}}$ of the source particle over a fixed time $t$,  treating each displacement as an effective velocity $\mathbf{u}_{\mathrm{source}} = (\Delta y_\mathrm{source}/t) \, \hat{\textbf{y}}$. We, thus, measure the conditional vorticity, $\langle \pmb{\omega}_{\mathrm{exp}} | \mathbf{u}_{\mathrm{source}} \rangle$, by partitioning the ensemble of source-particle velocities into bins of fixed magnitude, $|\mathbf{u}_{\mathrm{source}}|$. For each bin, we then compute the corresponding $\langle \pmb{\omega}_{\mathrm{exp}} \rangle$ in the surrounding fluid, thereby isolating the hydrodynamic response to a net Brownian displacement of well-defined magnitude over a fixed lag time. This conditional analysis reveals a clear linear relationship: larger source particle velocities produce proportionally stronger vorticity fields [Fig.~\ref{fig:Figure3}(b)]. These results show that vorticity is not solely a passive imprint of solvent fluctuations; thermally driven particle motion continually generates flow disturbances that remain hydrodynamically coupled across the suspension.

\paragraph{Phenomenological Model}---The experimentally observed ultraslow decay of $\langle \pmb{\omega}_{\mathrm{exp}}(r, t) \rangle$ thus signals the breakdown of the purely diffusive picture of vorticity as readily inferred from the ``classic'' unsteady Stokes equation. In a moderately dense suspension, the Stokeslet flow generated by the source particle displaces neighboring particles, whose own Stokeslet responses act back on the solvent and source particle. In the Stokes regime, this hydrodynamic back-coupling produces a history force whose kernel decays as a power law in time, generating long-lived hydrodynamic memory \cite{Franosch2011, Kheifets2014} that substantially reshapes the temporal evolution of vorticity. To capture this physics, we extend the unsteady Stokes description by incorporating a time-dependent memory kernel that reflects the stochastic, temporally correlated velocity fluctuations of the source particle within the moderately concentrated suspension~\cite{Batchelor1976, Delorme2016}.

Let us assume that vorticity is generated by a localized (point) source at the origin, arising from the stochastic velocity of the source particle. The vorticity transport equation then becomes:
\begin{equation}
\partial_t \pmb{ \omega } = \nu \nabla^2 \pmb{\omega} + 3\pi \nu \sigma \, \nabla \times \big(\mathbf{u}_\mathrm{source}(t) \, \delta(\mathbf{r}) \big),
\label{eq:unsteadySE}
\end{equation}
where the Stokes drag is $\mathbf{F} = 3\pi \mu \sigma  \mathbf{u}_\mathrm{source}$ (the friction coefficient is $\gamma = 3\pi \mu \sigma$) and the source term enters as a force density, $\mathbf{F}\delta(\mathbf{r})/\rho$. Using the Green's function for the three-dimensional heat equation, $G_0(\mathbf{r}, t) = \frac{1}{(4\pi\nu t)^{3/2}} \exp\!\left(-\frac{r^2}{4\nu t}\right)$, the vorticity field follows by convolution, 
\begin{equation}
    \pmb{\omega}(\mathbf{r},t) = \int_0^tdt'\int_{\mathbb{R}^3} d^3\mathbf{r}'G_0(\mathbf{r}- \mathbf{r}', |t-t'|)\mathbf{S}(\mathbf{r}',t'),
\end{equation}
with $\mathbf{S}(\mathbf{r},t) = 3\pi\nu\sigma \nabla \times (\mathbf{u}_{\mathrm{source}}(t) \delta (r))$. Working in cylindrical coordinates $(s, \phi, z)$, and aligning the source velocity along the axial direction, $\mathbf{u}_\mathrm{source} = u_z(t) \hat{\textbf{z}}$, the spatial integral can be performed (see Supplemental Note S2 \cite{supplemental}) to yield the azimuthal vorticity,
\begin{equation}
    \pmb{\omega}(\mathbf{r},t) = - 3\pi\nu\sigma \hat{\pmb{\phi}} \int_0^t u_z(t') \frac{\partial G_0(\mathbf{r},|t-t'|)}{\partial s} dt'.
\end{equation}

Because our experimental measurements are performed in moderately concentrated suspensions and report the ensemble-averaged vorticity field, $\langle\cdot \rangle$, we must average over all stochastic particle trajectories that produce a given net displacement $\Delta z$ over a lag time $t$. By the linearity of Eq.~\ref{eq:unsteadySE}, the averaging acts only on the stochastic source velocity, so the ensemble-averaged azimuthal vorticity conditioned on $\Delta z$ becomes 
\begin{equation}
    \langle \omega_\phi(\mathbf{r}, t) \rangle= -3\pi \nu \sigma \int_0^t \langle u_z(t')\rangle_{_{\Delta z}}  \frac{\partial G_0(\mathbf{r},|t-t'|)}{\partial s} dt'. 
    \label{eq:ensemble_vorticity}
\end{equation}
Here, $\langle u_z(t')\rangle_{_{\Delta z}}$ denotes the mean (conditional) source velocity at time $t'$ given that the particle's net displacement over $[0,t]$ equals $\Delta z$. Physically, it is the average instantaneous source-particle velocity at an intermediate time $t'$ conditioned on the particle reaching the prescribed endpoint $\Delta z$ at time $t$, i.e. a conditioned path average over stochastic trajectories connecting the initial and final positions. Eq.~\ref{eq:ensemble_vorticity} is the basis for quantitatively linking measured conditional displacements to the observed, slowly decaying vorticity.

To evaluate Eq.~\ref{eq:ensemble_vorticity}, we need to know the statistical properties of $u_z (t')$. As a phenomenological approximation, we model the source-particle velocity fluctuations as a zero-mean Gaussian process. Since there is no net drift, the unconditional mean of $u_z (t')$ must be zero. Hydrodynamic memory, however, is known to generate long-time velocity correlations arising from fluid inertia, as captured by generalized Langevin approaches featuring the Basset--Boussinesq memory kernel~\cite{guyon2015physical, goychuk2019fractional, lesnicki2016molecular,Delorme2016, mainardi2008fractional, szymanski2009elucidating}. Rather than solving the generalized Langevin equations, we adopt a phenomenological description in terms of fractional Brownian motion (fBM), which reproduces power-law temporal correlations in a compact stochastic framework \cite{guyon2015physical, goychuk2019fractional, lesnicki2016molecular,Delorme2016, mainardi2008fractional, szymanski2009elucidating}. In this description, the conditional mean velocity of a particle that undergoes a net displacement $\Delta z$ over time $t$ becomes $\langle u_z(t') \rangle_{_{\Delta z}} = \frac{H \Delta z}{t^{2H}} \big[t'^{2H-1} + (|t-t'|)^{2H-1}\big]$, where $H$ is the Hurst exponent characterizing the degree of temporal correlation (see Supplemental Note S2 \cite{supplemental}). Because the memory integral is causal, the retarded-time variable satisfies $0 \le t' \le t$ throughout, so $|t-t'|^{2H-1}$ in fBM reduces identically to $(t-t')^{2H-1}$ on the integration domain. We emphasize that, unlike the hydrodynamic memory kernel derived from first-principles fluid mechanics, which fixes the long-time decay of VACF to the universal $t^{-3/2}$ form, the fBM model provides a coarse-grained approximation allowing for a tunable correlation exponent, $H$. In the limiting case of ordinary Brownian motion ($H = 1/2$), this expression simplifies to the memoryless form $\langle \mathbf{u}_{\mathrm{source}}(t') \rangle_{_{\Delta z}} = (\Delta z / t) \hat{\mathbf{z}}$~\cite{Zwanzig2001}. Thus, the framework offers a tractable means of incorporating effective velocity correlations, consistent with the experimentally observed stochastic forcing of the source particles, into the description of vorticity generation in our system.

 Incorporating $\langle u_z(t') \rangle_{_{\Delta z}}$, the solution of Eq.~\ref{eq:ensemble_vorticity} can be written as,
\begin{equation}
    \langle \omega_\phi(\mathbf{r}, t) \rangle = -\frac{3\pi\nu \sigma H \Delta z}{t^{2H}}
    \frac{\partial \Psi_H(\mathbf{r}, t)}{\partial s},
    \label{eq:omega}
\end{equation}
where $\Psi_H(\mathbf{r}, t)$ encodes the hydrodynamic memory: $\Psi_H(\mathbf{r}, t) = \int_0^t dt'~ \![t'^{2H-1} + (t - t')^{2H-1}] G_0(\mathbf{r}, t - t')$. For $H < 1/2$, corresponding to \textit{anti-persistent} hydrodynamic memory, a classic feature of hydrodynamically correlated motion in complex fluids~\cite{Franosch2011, mandal2019persistent, Vinales2020}, the asymptotic form of Eq.~\ref{eq:omega} for $t \gg r^2/\nu$ reduces to (see Supplemental Note S2 \cite{supplemental})
\begin{equation}
    \langle \omega_\phi(s,z,t) \rangle
    \simeq
    \frac{3\pi\nu\sigma H(3 - 4H)\,\Gamma\!\left(\tfrac{3}{2} - 2H\right)}
    {(4\nu)^{2H}\pi^{3/2} r^{4-4H}} \frac{s}{r}
    \frac{\Delta z\,}{\,t^{2H}}.
    \label{eq:omega_asymptotic}
\end{equation}

In the 2D cross-section of the 3D volume probed experimentally, we have $s = r$. With the displacement of the source particles aligned along the $y$-axis, \textit{i.e.}, $ \Delta y_{\mathrm{source}} \,\hat{\mathbf{y}}$, the vorticity evolution reduces to 
\begin{equation}
    \langle \omega (r,t) \rangle \simeq K \frac{\Delta y_\mathrm{source}}{t^{2H}} \frac{1}{r^{4-4H}}, 
    \label{eq:omega_2D}
\end{equation}
where the constant $K = \frac{3\pi \nu \sigma\, H\, (3-4H)\, \Gamma\!\left(3/2-2H\right)} {\pi^{3/2} (4\nu)^{2H}}$ with $\Gamma(\mathscr{x})$ is the standard gamma function (see Supplemental Note S2 \cite{supplemental}). Remarkably, for $H = 0.40$, the model quantitatively reproduces both the experimentally observed temporal decay of vorticity [Fig.~\ref{fig:Figure3}(a)] and its spatial decay with $r$ (see Supplemental Fig.~S4 \cite{supplemental}). The exponent $H$ is not merely a fit parameter: within the fBM framework, anti-persistence ($H<1/2$) implies subdiffusive mean-squared displacement (MSD) scaling, $\langle \Delta r^2(t) \rangle \sim t^{2H}$, independently of the vorticity analysis. The short-time MSD measured from our experiments yields $H = 0.43 \pm 0.02$ (Supplemental Fig.~S5 \cite{supplemental}), consistent with $H = 0.40$ within experimental uncertainty. The \emph{effective} $\nu$, determined from fitting Eq.~\ref{eq:omega_2D}, is found to be $r$-dependent, but saturates at $\nu \sim 7.4 \times 10^{-6}\,\mathrm{m^2/s}$, consistent with expectations (End Matter for Materials and Methods and see Supplemental Fig.~S6 \cite{supplemental}). Although vorticity diffuses orders of magnitude faster than particle transport ($\sim 10^{-14}\,\mathrm{m^2\,s^{-1}}$; see Supplemental Fig.~S6 and Fig.~S7 \cite{supplemental}), hydrodynamic memory slows its decay, rendering the vorticity persistent and directly observable.

Finally, to establish the robustness of these findings, we performed experiments at a lower volume fraction $\Phi = 0.08$ and complementary overdamped Langevin dynamics simulations at $\Phi = 0.12$ without hydrodynamic interactions \cite{LAMMPS} (Supplemental Figs.~S3, S5, and S8 \cite{supplemental}). At $\Phi = 0.08$, the conditioned vorticity field retains the spatial structure and power-law temporal decay observed at $\Phi = 0.12$, confirming that the effect is robust across concentrations. In contrast, the Langevin dynamics simulations, which preserve Brownian statistics but remove hydrodynamic coupling, produce no systematic vortical response under the same conditioning procedure. Together, these results establish that Brownian motion alone is insufficient to sustain the observed vorticity. Rather, the results are consistent with Stokeslet-like flows generated by thermally driven particle motion, coupled through suspension hydrodynamics, giving rise to an effective power-law memory kernel in the effective forcing, driving the ultraslow vorticity decay reported here.

\paragraph{Discussion---}Anti-persistent Brownian motion ($H < 1/2$) is well established for isolated colloidal particles, arising from hydrodynamic memory associated with momentum diffusion in the solvent~\cite{Franosch2011, Vinales2020}. These studies demonstrate that even in equilibrium fluids, hydrodynamic coupling between particle motion and the surrounding liquid produces temporally correlated noise rather than white noise. For a single particle, however, vorticity is embedded implicitly in the noise kernel and decays on timescales set by $\tau_p$. In a moderately concentrated suspension, many-body hydrodynamic back-coupling can sustain and enhance the collective vortical response, extending its observable lifetime to timescales of order $0.01$--$1\,\mathrm{s}$, orders of magnitude beyond $\tau_p$, thereby making particle-resolved vorticity directly measurable.

Here, by directly measuring the conditioned vorticity field and interpreting it within a phenomenological framework that incorporates a distribution of stochastic forcing amplitudes, we find that the resulting vorticity decays far more slowly than predicted by the unforced unsteady Stokes response. The algebraic decay reflects the combined influence of anti-persistent fractional noise, hydrodynamic coupling among fluctuating particles, and diffusive momentum transport. The quantitative agreement between experiment and fractional-memory framework indicates that, even without externally applied forcing, moderately dense suspensions retain a long-lived hydrodynamic memory that manifests at the field level through persistent vorticity. What minimum strength of many-body coupling is required, and whether a corresponding lower bound in $\Phi$ exists, for the particle-resolved vortical response to be resolvable through single-particle tracking, remains an important open question.

This behavior suggests that moderately dense passive suspensions support collective, fluctuation-generated flow structures that cannot be captured by models assuming $\delta$-correlated forcing or Markovian dissipation. Such long-lived correlations may imply modifications to transport relations, such as deviations from the Stokes--Einstein relation~\cite{Barbhuiya2023, edmond2012decoupling, chen2006stokes, xu2009appearance, grober2023unconventional}, when memory effects are present. While analogous signatures of noise-supported vorticity have been reported in low-Reynolds-number systems ~\cite{dunkel2013fluid, riedel2005self, gautam2024harnessing, mukherjee2023intermittency}, where energy injection is explicit, our findings show that similar phenomenology may arise in equilibrium suspensions when hydrodynamic memory and multi-source stochastic forcing coexist. Thus, vorticity becomes not just a passive consequence of motion, but a persistent dynamical field encoding the collective memory of microscopic fluctuations.

\begin{acknowledgments}

\paragraph*{Acknowledgments---}
The authors thank Sankalp Nambiar, Adhip Agarwala, Prasanna Venkatesh B., and Anupam Gupta for discussions.  We gratefully acknowledge financial support from the Department of Science and Technology (Government of India), INSPIRE Fellowship IF200274 (NHB) and the Anusandhan National Research Foundation (ANRF) under the Advanced Research Grant (ARG) program [Grant No. ANRF/ARG/2025/002467/PS] (CKM) and IIT Gandhinagar (CKM). T.T. acknowledges support from the National Research Foundation of Korea (NRF-RS-2025-00573354), the InnoCORE Bio-MAX program, and the U.S. Office of Naval Research  (N00014-26-1-2401).
\end{acknowledgments}


\bibliography{reference}

@misc{supplemental,
  title = {See Supplemental Material [{URL}] for details on streamline rendering, derivation of the vorticity transport equation with fractional {B}rownian memory, asymptotic analysis, and the tracer-based proxy for the fluid vorticity field, along with additional figures showing the individual trajectories conditioned on a fixed displacement, spatio-temporal evolution of vorticity field, robustness of the conditional vorticity signal, spatial profile of vorticity, short-time mean-squared displacements, spatial profile of fitted kinematic viscosity, Single-particle validation of {B}rownian dynamics, hydrodynamic field and vorticity at a lower volume fractions ({$\Phi = 0.08$}), particle size distribution, pair correlation function, Fax\'{e}n correction error analysis and captions for supplemental videos of confocal microscopy and vorticity field dynamics},
  note = {which includes Refs.~\cite{guyon2015physical, Delorme2016}}
}

@Article{LAMMPS,
  author = "A. P. Thompson and H. M. Aktulga and R. Berger and 
     D. S. Bolintineanu and W. M. Brown and P. S. Crozier and
     P. J. in 't Veld and A. Kohlmeyer and S. G. Moore and T. D. Nguyen and
     R. Shan and M. J. Stevens and J. Tranchida and C. Trott and S. J. Plimpton",
  title = "{LAMMPS} - a flexible simulation tool for
     particle-based materials modeling at the 
     atomic, meso, and continuum scales",
  journal = "Comp. Phys. Comm.",
  volume =  "271",
  pages =   "108171",
  year =    "2022",
  doi = "10.1016/j.cpc.2021.108171"
}

@book{batchelor2000introduction,
  title={An introduction to fluid dynamics},
  author={Batchelor, George Keith},
  year={2000},
  publisher={Cambridge university press}
}

@book{frisch1995turbulence,
  title={Turbulence: the legacy of AN Kolmogorov},
  author={Frisch, Uriel and Kolmogorov, A N},
  year={1995},
  publisher={Cambridge university press}
}

@article{sreenivasan1999fluid,
  title={Fluid turbulence},
  author={Sreenivasan, Katepalli R},
  journal={Rev. Mod. Phys.},
  volume={71},
  number={2},
  pages={S383},
  year={1999}
}

@article{eyink2006onsager,
  title={Onsager and the theory of hydrodynamic turbulence},
  author={Eyink, Gregory L and Sreenivasan, Katepalli R},
  journal={Rev. Mod. Phys.},
  volume={78},
  number={1},
  pages={87},
  year={2006}
}

@article{holland1984dynamics,
  title={Dynamics of the oceanic general circulation: the potential vorticity field},
  author={Holland, WR and Keffer, T and Rhines, PB},
  journal={Nature},
  volume={308},
  number={5961},
  pages={698},
  year={1984}
}

@article{buaria2020self,
  title={Self-attenuation of extreme events in Navier--Stokes turbulence},
  author={Buaria, Dhawal and Pumir, Alain and Bodenschatz, Eberhard},
  journal={Nat. Commun.},
  volume={11},
  number={1},
  pages={5852},
  year={2020}
}

@article{buaria2024twisting,
  title={Twisting vortex lines regularize Navier-Stokes turbulence},
  author={Buaria, Dhawal and Lawson, John M and Wilczek, Michael},
  journal={Sci. Adv.},
  volume={10},
  number={37},
  pages={eado1969},
  year={2024}
}

@article{de2024pattern,
  title={Pattern formation by turbulent cascades},
  author={de Wit, Xander M and Fruchart, Michel and Khain, Tali and Toschi, Federico and Vitelli, Vincenzo},
  journal={Nature},
  volume={627},
  number={8004},
  pages={515},
  year={2024}
}

@book{guyon2015physical,
  title={Physical hydrodynamics},
  author={Guyon, Etienne and Hulin, Jean Pierre and Petit, Luc and Mitescu, Catalin D},
  year={2015},
  publisher={Oxford university press}
}

@book{Happel1983,
  author    = {Happel, John and Brenner, Howard},
  title     = {Low {Reynolds} Number Hydrodynamics},
  year      = {1983},
  publisher = {Martinus Nijhoff Publishers},
  address   = {The Hague},
  isbn      = {9789024728770}
}

@article{cuevas2006vorticity,
  title={Vorticity generation in creeping flow past a magnetic obstacle},
  author={Cuevas, S and Smolentsev, S and Abdou, M},
  journal={Phys. Rev. E},
  volume={74},
  number={5},
  pages={056301},
  year={2006}
}

@book{Kim2005,
  author    = {Kim, Sangtae and Karrila, Seppo J.},
  title     = {Microhydrodynamics: Principles and Selected Applications},
  year      = {2005},
  publisher = {Dover Publications},
  address   = {Mineola, NY},
  isbn      = {9780486442198}
}

@article{Batchelor1976,
  author    = {Batchelor, G. K.},
  title     = {Brownian diffusion of particles with hydrodynamic interaction},
  journal   = {J. Fluid Mech.},
  year      = {1976},
  volume    = {74},
  number    = {1},
  pages     = {1}
}

@book{landau2013fluid,
  title={Fluid Mechanics: Volume 6},
  author={Landau, L.D. and Lifshitz, E.M.},
  number={v. 6},
  isbn={9781483140506},
  year={2013},
  publisher={Pergamon}
}

@article{keblinski2006hydrodynamic,
  title={Hydrodynamic field around a Brownian particle},
  author={Keblinski, P and Thomin, J},
  journal={Phys. Rev. E},
  volume={73},
  number={1},
  pages={010502},
  year={2006}
}

@article{fares2024observation,
  title={Observation of Brownian elastohydrodynamic forces acting on confined soft colloids},
  author={Fares, Nicolas and Lavaud, Maxime and Zhang, Zaicheng and Jha, Aditya and Amarouchene, Yacine and Salez, Thomas},
  journal={Proc. Natl. Acad. Sci. USA},
  volume={121},
  number={42},
  pages={e2411956121},
  year={2024}
}

@article{lukic2005direct,
  title={Direct observation of nondiffusive motion of a Brownian particle},
  author={Luki{\'c}, B and Jeney, S and Tischer, C and Kulik, AJ and Forr{\'o}, L and Florin, E-L},
  journal={Phys. Rev. Lett.},
  volume={95},
  number={16},
  pages={160601},
  year={2005}
}

@article{Franosch2011,
  author    = {Franosch, Thomas and Grimm, Matthias and Belushkin, Maxim and Mor, Flavio M. and Foffi, Giuseppe and Forr{\'o}, L{\'a}szl{\'o} and Jeney, Sylvia},
  title     = {Resonances arising from hydrodynamic memory in {Brownian} motion},
  journal   = {Nature},
  year      = {2011},
  volume    = {478},
  number    = {7367},
  pages     = {85},
  month     = {Oct}
}

@article{huang2011direct,
  title={Direct observation of the full transition from ballistic to diffusive Brownian motion in a liquid},
  author={Huang, Rongxin and Chavez, Isaac and Taute, Katja M and Luki{\'c}, Branimir and Jeney, Sylvia and Raizen, Mark G and Florin, Ernst-Ludwig},
  journal={Nat. Phys.},
  volume={7},
  number={7},
  pages={576},
  year={2011}
}

@article{Kheifets2014,
  author    = {Kheifets, Simon and Simha, Akarsh and Melin, Kevin and Li, Tongcang and Raizen, Mark G.},
  title     = {Observation of {Brownian} motion in liquids at short times: instantaneous velocity and memory loss},
  journal   = {Science},
  year      = {2014},
  volume    = {343},
  number    = {6178},
  pages     = {1493},
  month     = {March}
}

@article{huang2015effect,
  title={Effect of interfaces on the nearby Brownian motion},
  author={Huang, Kai and Szlufarska, Izabela},
  journal={Nat. Commun.},
  volume={6},
  number={1},
  pages={8558},
  year={2015}
}

@article{mandal2019persistent,
  title={Persistent anti-correlations in Brownian dynamics simulations of dense colloidal suspensions revealed by noise suppression},
  author={Mandal, Suvendu and Schrack, Lukas and L{\"o}wen, Hartmut and Sperl, Matthias and Franosch, Thomas},
  journal={Phys. Rev. Lett.},
  volume={123},
  number={16},
  pages={168001},
  year={2019}
}

@article{Delorme2016,
  author    = {Delorme, Mathieu and Wiese, Kay J{\"o}rg},
  title     = {Maximum of a Fractional {Brownian} Motion: Analytic Results from Perturbation Theory},
  journal   = {Phys. Rev. Lett.},
  year      = {2016},
  volume    = {115},
  number    = {21},
  pages     = {210601}
}

@article{mainardi2008fractional,
  title={The fractional Langevin equation: Brownian motion revisited},
  author={Mainardi, Francesco and Pironi, Paolo},
  journal={arXiv preprint arXiv:0806.1010},
  year={2008}
}

@article{szymanski2009elucidating,
  title={Elucidating the origin of anomalous diffusion in crowded fluids},
  author={Szymanski, Jedrzej and Weiss, Matthias},
  journal={Phys. Rev. Lett.},
  volume={103},
  number={3},
  pages={038102},
  year={2009},
  publisher={APS}
}

@book{Zwanzig2001,
  author    = {Zwanzig, Robert},
  title     = {Nonequilibrium Statistical Mechanics},
  year      = {2001},
  publisher = {Oxford University Press},
  address   = {Oxford}
}

@article{Vinales2020,
  author    = {Vi{\~n}ales, A. D. and Camuyrano, M. and Paissan, G. H.},
  title     = {Oscillations and negative velocity autocorrelation emerging from a {Brownian} particle model with hydrodynamic interactions},
  journal   = {Phys. Rev. E},
  year      = {2020},
  volume    = {101},
  number    = {5},
  pages     = {052140}
}

@article{Crocker1996,
  author    = {Crocker, John C. and Grier, David G.},
  title     = {Methods of digital video microscopy for colloidal studies},
  journal   = {J. Colloid Interface Sci.},
  year      = {1996},
  volume    = {179},
  number    = {1},
  pages     = {298}
}

@article{Segre2001,
  author    = {Segr{\'e}, P. N. and Liu, F. and Umbanhowar, P. and Weitz, D. A.},
  title     = {An effective gravitational temperature for sedimentation},
  journal   = {Nature},
  year      = {2001},
  volume    = {409},
  number    = {6820},
  pages     = {594}
}

@article{leunissen2005ionic,
  title={Ionic colloidal crystals of oppositely charged particles},
  author={Leunissen, Mirjam E and Christova, Christina G and Hynninen, Antti-Pekka and Royall, C Patrick and Campbell, Andrew I and Imhof, Arnout and Dijkstra, Marjolein and Van Roij, Rene and Van Blaaderen, Alfons},
  journal={Nature},
  volume={437},
  number={7056},
  pages={235},
  year={2005}
}

@phdthesis{leunissen2007manipulating,
  title={Manipulating colloids with charges and electric fields},
  author={Leunissen, Mirjam E},
  year={2007},
  school={Utrecht University}
}

@article{riedel2005self,
  title={A self-organized vortex array of hydrodynamically entrained sperm cells},
  author={Riedel, Ingmar H and Kruse, Karsten and Howard, Jonathon},
  journal={Science},
  volume={309},
  number={5732},
  pages={300},
  year={2005}
}

@article{dunkel2013fluid,
  title={Fluid dynamics of bacterial turbulence},
  author={Dunkel, J{"o}rn and Heidenreich, Sebastian and Drescher, Knut and Wensink, Henricus H and B{"a}r, Markus and Goldstein, Raymond E},
  journal={Phys. Rev. Lett.},
  volume={110},
  number={22},
  pages={228102},
  year={2013}
}

@article{mukherjee2023intermittency,
  title={Intermittency, fluctuations and maximal chaos in an emergent universal state of active turbulence},
  author={Mukherjee, Siddhartha and Singh, Rahul K and James, Martin and Ray, Samriddhi Sankar},
  journal={Nat. Phys.},
  volume={19},
  number={6},
  pages={891},
  year={2023}
}

@article{gautam2024harnessing,
  title={Harnessing density to control the duration of intermittent L{\'e}vy walks in bacterial turbulence},
  author={Gautam, Dhananjay and Meena, Hemlata and Matheshwaran, Saravanan and Chandran, Sivasurender},
  journal={Phys. Rev. E},
  volume={110},
  number={1},
  pages={L012601},
  year={2024}
}

@article{Barbhuiya2023,
  author    = {Barbhuiya, Noman Hanif and Yodh, Arjun G. and Mishra, Chandan K.},
  title     = {Direction-dependent dynamics of colloidal particle pairs and the {Stokes-Einstein} relation in quasi-two-dimensional fluids},
  journal   = {Nat. Commun.},
  year      = {2023},
  volume    = {14},
  number    = {1},
  pages     = {4991},

}

@article{edmond2012decoupling,
  title={Decoupling of rotational and translational diffusion in supercooled colloidal fluids},
  author={Edmond, Kazem V and Elsesser, Mark T and Hunter, Gary L and Pine, David J and Weeks, Eric R},
  journal={Proceedings of the National Academy of Sciences},
  volume={109},
  number={44},
  pages={17891--17896},
  year={2012},
  publisher={National Academy of Sciences}
}

@article{chen2006stokes,
  title={Stokes-Einstein relation in supercooled aqueous solutions of glycerol},
  author={Chen, Bo and Sigmund, Eric E and Halperin, WP},
  journal={Physical review letters},
  volume={96},
  number={14},
  pages={145502},
  year={2006},
  publisher={APS}
}

@article{xu2009appearance,
  title={Appearance of a fractional Stokes--Einstein relation in water and a structural interpretation of its onset},
  author={Xu, Limei and Mallamace, Francesco and Yan, Zhenyu and Starr, Francis W and Buldyrev, Sergey V and Eugene Stanley, H},
  journal={Nature Physics},
  volume={5},
  number={8},
  pages={565--569},
  year={2009},
  publisher={Nature Publishing Group UK London}
}

@article{grober2023unconventional,
  title={Unconventional colloidal aggregation in chiral bacterial baths},
  author={Grober, Daniel and Palaia, Ivan and U{\c{c}}ar, Mehmet Can and Hannezo, Edouard and {\v{S}}ari{\'c}, An{\dj}ela and Palacci, J{\'e}r{\'e}mie},
  journal={Nature Physics},
  volume={19},
  number={11},
  pages={1680--1688},
  year={2023},
  publisher={Nature Publishing Group UK London}
}

@article{lesnicki2016molecular,
  title={Molecular hydrodynamics from memory kernels},
  author={Lesnicki, Dominika and Vuilleumier, Rodolphe and Carof, Antoine and Rotenberg, Benjamin},
  journal={Phys. Rev. Lett.},
  volume={116},
  number={14},
  pages={147804},
  year={2016},
  publisher={APS}
}

@article{goychuk2019fractional,
  title={Fractional hydrodynamic memory and superdiffusion in tilted washboard potentials},
  author={Goychuk, Igor},
  journal={Phys. Rev. Lett.},
  volume={123},
  number={18},
  pages={180603},
  year={2019},
  publisher={APS}
}


\section*{End Matter}

\subsection*{Materials and Methods}

\paragraph{\textbf{Sample Preparation:}}
The colloidal suspension consisted of poly(methyl methacrylate) (PMMA) spheres of diameter $\sigma = 2.2~\mu\text{m}$ (mass density $\rho \approx 1.19~\text{g/mL}$, refractive index $n \approx 1.492$; see Supplemental Fig.~S9 for particle size distribution \cite{supplemental}), sterically stabilized by a poly(12-hydroxystearic acid) (PHSA) layer and fluorescently labeled with nitrobenzoxadiazole (NBD). To minimize gravitational effects and ensure a spatially homogeneous thermally equilibrated system, the particles were dispersed in a binary solvent matched in both density and refractive index. The solvent comprised cyclohexyl bromide (CHB; $\rho \approx 1.32~\text{g/mL}$, $n \approx 1.495$) and decalin ($\rho \approx 0.88~\text{g/mL}$, $n \approx 1.474$) mixed in a 65.5:34.5 (v/v) ratio, yielding near-perfect density matching with the PMMA particles~\cite{leunissen2005ionic, leunissen2007manipulating}. Residual electrostatic interactions were screened by saturating the decalin phase with tetra-\emph{n}-butylammonium bromide (TBAB), resulting in nearly hard-sphere behavior~\cite{leunissen2005ionic, leunissen2007manipulating}. Samples were prepared at particle volume fractions $\Phi \approx 0.12$ and $\Phi \approx 0.08$ (see Supplemental Fig.~S10 for their pair-correlation function \cite{supplemental}) and sealed to prevent solvent evaporation during imaging. Assuming that the dynamic viscosity of the suspension is comparable to that of the solvent (decalin), for which manufacturer-reported values are available, we estimate the kinematic viscosity of the index- and density-matched colloidal suspension and it turns out to be $\nu \sim 10^{-6}~\mathrm{m^2/s}$.

\paragraph{\textbf{Optical Video Microscopy:}}
Imaging was performed using a Leica Stellaris 8 confocal microscope equipped with a 63$\times$, 1.4 NA oil-immersion objective. The NBD-labeled particles were excited with a 488~nm laser line. Time-series data were recorded from a region of interest positioned well away from the chamber walls [Fig.~\ref{fig:Figure1}(a)]. To capture the full three-dimensional particle dynamics, volumetric stacks spanning $\sim 92 \times 23 \times 23~\mu\mathrm{m}^3$ were acquired at an acquisition rate of 0.58~fps, with successive $z$-slices separated by 200~nm (see video~S1 \cite{supplemental}). To probe faster dynamics with higher temporal resolution, a single fixed mid-plane of size $\sim 92 \times 92~\mu\mathrm{m}^2$ was imaged at 21.54~fps [Fig.~\ref{fig:Figure1}(a)]. This high-speed acquisition was crucial for resolving the temporal decay of vorticity fields. Particle trajectories were extracted from both 2D and 3D image series using standard tracking algorithms~\cite{Crocker1996}, and subsequent analyses were performed using in-house-developed Python codes.

\paragraph{\textbf{Constructing Velocity Field Map:}}
To construct the velocity field map corresponding to a given lag time $t$, we proceeded as follows. Over this lag each particle undergoes a displacement $\Delta \mathbf{r}_i(t_0,t) = \mathbf{r}_i(t_0 + t) - \mathbf{r}_i(t_0)$ with an average velocity $\mathbf{u}_i(t_0,t) = \Delta \mathbf{r}_i(t_0,t)/t$, where $t_0$ is an arbitrary initial time. Each particle is sequentially chosen as the \textit{source particle}. For a given source, its velocity vector is rotated by an angle $\theta$ such that it aligns along a fixed reference direction (the $z$-axis for the 3D data and the $y$-axis for the 2D data). The velocities of all surrounding particles are rotated by the same angle $\theta$, preserving their relative orientations with respect to the source. An ensemble average is then performed over all particles, together with an averaging over all initial times $t_0$. The averaged velocity in each grid element around the source particle yields the velocity field map corresponding to the lag time $t$. For the 3D data, the volume surrounding each source particle is divided into cubic bins of size $0.25\sigma$, while for the 2D data, the plane is partitioned into square bins of size $0.41\,\sigma$. The resulting velocity field maps are shown in Fig.~\ref{fig:Figure2}.

\paragraph{\textbf{Calculating the Vorticity Field:}}
All the vorticity analysis reported in the text was performed using the 2D data, chosen for the higher frame rate of image acquisition and its statistical robustness (compared to 3D data). From the constructed 2D velocity field map, $\mathbf{u}(x, y)$, we calculated the vorticity, $\pmb{\omega}(x, y)$, defined as the curl of the velocity field. For a two-dimensional flow in the $x$--$y$ plane, the vorticity vector is perpendicular to the plane (along the $z$-direction), and its magnitude is given by
$$|\omega_z| = |\left( \nabla \times \mathbf{u} \right)_z| = \left|\frac{\partial u_y}{\partial x} - \frac{\partial u_x}{\partial y}\right|.$$
Since the velocity field is reconstructed from tracer-particle displacements rather than direct measurements on the fluid, the quantity $|\omega_z|$ is a tracer-based proxy for the true solvent vorticity. The validity of this proxy rests on the Fax\'{e}n theorem: in the far field ($ r > \sigma$), the Fax\'{e}n corrections to the tracer velocity are of order $\mathcal{O}(\sigma/r)^2$ and are negligibly small, so tracer displacements can be considered to follow the local solvent velocity faithfully (Supplemental Note S3 \& Fig. S11 \cite{supplemental}) \cite{Kim2005}. This quantity was computed numerically for each bin in the discretized grid using a central finite-difference scheme on the averaged velocity data. To improve the statistical accuracy of the vorticity map, we exploited the expected cylindrical symmetry of the flow field about the $y$-axis. This symmetry implies that the vorticity magnitudes at coordinates $(x, y)$ and $(-x, y)$ should be equivalent. However, during sampling, the number of velocity vectors contributing to each bin can differ between these symmetric locations. To account for this, we symmetrized the vorticity by averaging the magnitudes from bins at $(x, y)$ and $(-x, y)$, weighting each contribution according to the number of velocity vectors in the corresponding bin. This procedure ensures that bins with more data have proportionally greater influence on the final, symmetric vorticity field.

\paragraph{\textbf{Overdamped Langevin Dynamics Simulations:}}
To test whether the conditioned vortical signal can arise without hydrodynamic interactions, we performed four independent overdamped Langevin dynamics simulations in LAMMPS \cite{LAMMPS} for $N = 4648$ spheres of diameter $\sigma = 2.2~\mu\mathrm{m}$ at $\Phi = 0.12$ in a cubic periodic box of side length $60 \sigma$. Particle interactions were modeled by a purely repulsive Weeks--Chandler--Andersen-like potential with $\epsilon = 50\,k_B T$ and cutoff $r_c = 2^{1/6}\sigma$. Four independent replicas were used, each initialized from a distinct configuration with independent random seeds for the initial velocities and Langevin noise, to verify reproducibility of the results.

The equations of motion were integrated using velocity-Verlet scheme with a Langevin thermostat, which models the solvent implicitly and excludes hydrodynamic interactions by construction. The damping time and particle mass were chosen such that the free-particle diffusivity matched the experimental value $D_0 = 0.01\,\sigma^2/\mathrm{s}$, and the trajectories were sampled at the experimental frame rate of $21$~fps using $500$ integration steps per frame. Following an initial overlap-removal stage with a reduced timestep, each simulation was equilibrated for $200$~s, after which a production run for $1200$~s was collected. The identical conditioning and field-reconstruction procedure used for the experimental data was then applied to these trajectories, and the results are shown in Supplemental Figs.~S3(e) and S5.

\end{document}